\documentclass{ws-p8-50x6-00}
\newcommand{\bq}{\begin{equation}}
\newcommand{\ba}{\begin{eqnarray}}
\newcommand{\eq}{\end{equation}}
\newcommand{\ea}{\end{eqnarray}}

\def\c{\raisebox{.4ex}{$\chi$}}
\def\d{\delta}
\def\j{\psi}

\def\f{\phi}
\def\l{\lambda}
\def\r{\rho}

\def\cc{{\cal C}} 
\def\cd{{\cal D}}
\def\ci{{\cal I}}

\def\F{\Phi}
\def\J{\Psi}
\def\S{\Sigma}
\def\sq{{\lower.2ex\hbox{\large$\partial_{\mu} \partial^{\mu}$}}}
\begin{document}

\title{Initial Value Problems in Quantum Field Theory  \\
and the Feynman Path Integral
Representation of Schwinger's Closed Time Path Formalism}

\author{Fred Cooper}

\address{Theoretical Division, Los Alamos National Labs\\
Los Alamos NM 87545 \\and \\ Physics Department, Boston College \\
Chestnut Hill, MA 02167-3811
\\E-mail: fcooper@lanl.gov}


\maketitle

\abstracts{
Schwinger's Closed-Time Path (CTP) formalism is an elegant way to insure
causality for initial value problems in Quantum Field Theory. Feynman's Path Integral on the other
hand is much more amenable than Schwinger's differential approach (related to the Schwinger
Dyson equations) for non-perturbative (in coupling constant) expansions such
as the large-N expansion. By marrying the CTP formalism with a large-N expansion of
 Feynman's Path Integral approach, we are for the first time able  to study the dynamics of phase
transitions in quantum field theory settings. We  review the Feynman Path Integral
representation for the generating functional for the Green's functions described by an initial
density matrix.   We then show that the large-N expansion for the path integral forms a natural
non-perturbative framework for discussing phase transitions in quantum field theory as well as for
giving a space time description of a heavy ion collision.  We  review results for a time evolving
chiral phase transition and the degradation of an electric field due to pair production from
strong electric fields. }

\section{Introduction}
In this talk I would like to review work done on initial value problems over the
past 8 years and explain why  large-$N$ expansions
coupled with Schwinger's closed time formulation for the Green's functions is a 
good starting point for understanding the dynamics of many systems.  One of the
technical issues in doing initial value problems in Quantum Field Theory is to insure causality
of the update equations. This problem is completely solved by using the Closed Time Path
(CTP) Green's functions of the Schwinger-Mahanthappa-Keldysh formalism. On the other
hand, we want to have a starting point for calculations which allows us to 
encapsulate non-perturbative phenomena such as the phase structure of strongly-
interacting systems.  The Path Integral when expanded
about self-consistently determined mean fields does the required bookkeeping for
doing systematic expansions in non-perturbative (in coupling constant) domains.  

\section{Path Integral Formulation of Initial Value Problems-CTP formalism}
 The  preservation of causality is guaranteeed by the
 CTP formalism which was first discussed by Schwinger \cite{ref:Schwinger} in the context of
quantum Brownian motion . The application to field theory was by Mahanthappa et. al.  and
amplification of these ideas to quantum transport theory \cite{ref:SchKel} was elaborated and
developed  by Keldysh. The
original formulation was in a Schwinger Dyson
equation formulation. This was later  reformulated \cite{ref:Zhou} as  a Path Integral  by  Zhou
et. al, R. D.Jordan and Calzetta and Hu. 
The Green's functions that occur in the CTP formalism are also the ones that
naturally occur in a study of inclusive particle production.

 The generating
functional for the Green's functions is  obtained by considering the  ``in" state
matrix element in the presence of external sources, and 
 inserting a complete set of late time  $t'$ states. In this way one can express the
original initial time matrix element as a product of transition matrix elements
from $0$ to $t'$ and the time reversed (complex conjugate) matrix element from
$t'$ to $0$. Since each term in this product is a transition matrix element of
the usual or time reversed kind, standard path integral representations for each
may be introduced. 
\ba 
e^{i W[J_+,J_-]}&=&Z_{in}[J_+,J_-] \equiv \int [\cd \J]\langle in \vert \j 
\rangle_{J_-} \ \langle \j  \vert in \rangle_{J_+} \qquad\qquad\nonumber \\
 &=&
\int [\cd \J] \langle in \vert {\cal {T}}^* exp \left[-i \int_{0}^{t^{\prime}} dt 
J_{-} (t) \phi (t) \right] \vert \J ,t'\rangle \times \nonumber\\
 & & \qquad\langle
\J ,t' \vert {\cal {T}} \exp \left[ i \int_{0}^{t^{\prime}} dt J_+ (t) \phi (t)\right]
\vert in \rangle \label{eqCTP}  
\ea 
so that, for example, 
\ba 
&&{\d^4 W_{in}[J_+,J_-]
\over {\d J_+(t_3) \d J_+(t_4) \d J_-(t_1) \d J_-(t_2)}}\bigg\vert_{J_+=J_-0} \nonumber \\
&& =  <in |
 {\cal{T}}^* \{  \phi(t_1) \phi(t_2) \}  {\cal{T}} \{
\phi(t_3)  \phi(t_4) \}|in> 
\ea
 Here
$\phi(t) =
\phi(x,t)$ and we are supressing the coordinate dependence and the integration over the spatial
volume in what follows for notational simplicity.  Since
the time ordering in eq. (\ref{eqCTP}) is forward (denoted by $\cal{T}$) along the
time path from $0$ to $t'$ in the second transition matrix element, but backward
(denoted by ${\cal {T}}^*$) along the path from $t'$ to $0$ in the first matrix
element. Thus the name: {\em{ closed time path generating functional}}.  If we
deform the backward and forward directed segments of the path slightly in
opposite directions in the complex $t$ plane, the symbol ${\cal {T}}_{\cc}$ may be
introduced for path ordering along the full closed time contour, $\cc$.

Following Calzetta and Hu we
introduce the path integral representation for each transition matrix 
element in eq. (\ref{eqCTP})and obtain 
\ba
Z \left[ J_{+}, J_{-},\r \right] &=& \int [\cd \varphi]  [\cd\varphi^{\prime}] \ \langle \varphi \vert \r \vert \varphi^{\prime} \rangle \int [\cd \j]
\int_{\varphi}^{\j} [\cd \phi_+]\int_{\varphi^{\prime}}^{\j } [\cd \phi_-]\ 
\times \nonumber \\
& & \exp \left[i\int_{0}^{\infty} dt  \left(\ L[\phi_{+}] - L[\phi_{-}]+J_+\phi_+  - J_-  \phi_-  \right) \right]\ . \nonumber \\
\label{eqCTPgen} 
\ea

The double path integral over the fields $\phi_+$ and $\phi_-$ in (\ref{eqCTPgen})
suggests that we introduce a two component contravariant vector of field
variables by \ba \phi^a = \left(\begin{array}{c} \phi_+ \\ \phi_- \end{array}\right)\ ;
\qquad a=1,2
\ea
One raises and lowers indices in this vector space with a $2 \times 2$ matrix with
indefinite signature, namely \bq c_{ab} = diag \ (+1,-1) = c^{ab} \label{metr}
\eq so that, for example  \bq J^ac_{ab}\F^b = J_+ \phi_+ - J_-\phi_-\ .  \eq 
The correlation functions are  $2\times 2$ matrices. The components of the 
two point function is \bq G^{ab}(t,t') = {\d^{2} W \over \d
J_{a}(t) \d J_{b}(t')} \bigg\vert_{J =0}\ . \eq Explicitly, the components of
this $2\times 2$ matrix are   \ba G^{21}(t,t') &\equiv& G_> (t,t') = \ i{\rm
Tr}\{\r\ \F(t) \overline\F(t') \}_{con}\ ,\nonumber \\ G^{12}(t,t') &\equiv& G_<
(t,t') = \pm i{\rm Tr}\{\r\ \overline\F(t') \F(t) \}_{con} \\ G_F=G^{11}(t,t') &=& 
i{\rm Tr}\left\{\r\ {\cal {T}}[\F (t) \overline\F(t') ] \right\}_{con} = \theta (t,t') G_>
(t,t') + \theta (t',t) G_< (t,t')\nonumber \\G_{F^*}= G^{22}(t,t') &=&  i{\rm Tr}\left\{\r\
{\cal {T}}^* [\F (t) \overline\F(t')] \right\}_{con}= \theta (t',t) G_> (t,t') + \theta (t,t')
G_< (t,t') \nonumber  \label{matr} \ea   
The $\pm$ refer to Bose ($+$)or Fermi $(-)$. $\bar{\Phi} = \Phi^{\dag}$ for
charged Bosons and $\bar{\Psi} = \Psi^{\dag} \gamma^0$ for Dirac Fermions. 
An alternative generating functional 
\cite{causal}
 uses the Complex Path Ordered Form:
\ba Z_{\cc} \left[ J, \r \right] &= & {\rm Tr} \left\{ \r  \left(
{\cal {T}}_{\cc} \exp\left[i \int_{\cc} dt J (t) \phi (t) \right] \right)
\right\}\qquad\qquad\qquad\nonumber \\ &=& \int [\cd \varphi^1] \int [\cd \varphi^2]\
\langle \varphi^1 \vert \r \vert \varphi^2\ \rangle \int_{\varphi^1}^{\varphi^2} [\cd \phi] \exp
\left[i\int_{\cc} dt  \,\left( L[\phi] + J\phi\right) \right]\ . \label{Zfin}
\nonumber \\
 \ea

This is identical in structure to the usual
expression for the generating functional in the more familiar in-out formalism,
The only difference is that path ordering according to the complex time contour
$\cc$ replaces the ordinary time ordering prescription.
The propagator is now written in the equivalent form:
 \ba G(t,t')
&=&\Theta_{\cc} (t,t') G_>(t,t') + \Theta_{\cc} (t',t) G_<(t,t')
\label{CTPg} \ea where $\Theta_{\cc}$ is the CTP complex contour ordered theta
function (see \cite{ref:Zhou} \cite{causal} ).  

 With this
definition of $G(t,t')$, the Feynman rules are the
ordinary ones, and matrix indices are not required.
As we have shown previously \cite{causal} this complex contour form allows one to 
give simple rules for constructing arbitrary graphs which preserve causality.

\subsection{Inclusive Dilepton Production}
Using LSZ reductions formula, we can write the expression for the inclusive production
of dileptons in a form which relates this quantity to an off mass shell Green's function
exactly of the form given by Schwinger's CTP generating functional.  One finds \cite{dilep}:
\begin{eqnarray}
&&{E_k \over m} {E_k'\over m} {d^6N \over [d^3k] [d^3k']}(k,k';s,s')=  \nonumber \\
&&\int d^4 x_1
d^4 x_2 d^4 x_3 d^4 x_4  e^{ik(x_2-x_4)}
 e^{ik'(x_1-x_3)} 
\bar {v}_{k',s'}{\cal{D}}_{x_3} \bar{u} _{k,s}{\cal{D}}_{x_2} \times \nonumber \\
&& \,  _{in} < P_1 P_2 |
 {\cal{T}}^* \{  \Psi(x_3) \bar{\Psi}(x_4) \}  {\cal{T}} \{
\Psi(x_2) \bar{\Psi}(x_1)  \}| P_1 P_2 >_ {in} \times \nonumber \\
&& \bar{ \cal{D}}_{x_4}u_{k,s} 
\bar{\cal{D}}_{x_1}v_{k',s'} \nonumber  \label{eq:dilep2} \\
\end{eqnarray}
Here $P_1$ and $P_2$ represent the two colliding particles at $t=t_0$.

\section{Large $N$}
Here we derive the causal evolution equation to order $1/N$ in the large $N$ expansion
\cite{largeN}
\cite{cgs}  for the
$O(N)$ model. The details are in reference 8. 
Following  Coleman, Jackiw and Politzer we write the Lagrangian as:
\begin{eqnarray}
\tilde L_{cl}[\Phi,\chi] &&=
-\frac{1}{2} \Phi_i (\sq+ \chi) 
\Phi_i +
{N\over \lambda} \chi\left({\chi \over 2}  +  \mu^2
\right)\nonumber\\ 
\label{lag}
\end{eqnarray}
where $i = 1, \ldots , N$ and $\chi$ is given by:
\begin{equation} 
\chi  = -\mu^2 + {\lambda\over 2N} \Phi_{i} \Phi_{i}~,  
\label{chi}
\end{equation}
If  $\mu^2 > 0$, spontaneous symmetry breaking at the classical level. 
The Generating functional for all Graphs is given by:
\begin{equation}
Z[j,K] = \int d\phi d\chi \exp \{i S[ \phi,\chi]+  i \int [j \phi + K \chi] \}
\end{equation}
Perform the Gaussian integral over the field $\phi$
\begin{equation}
Z[j,K]= \int d \chi \exp  \{i N S_{eff} [\chi, j,K] \}
\end{equation}
\[
 S_{eff} =\int dx \{ {1 \over 2} j G^{-1}[\chi] j +  K \chi +{1\over \lambda}
\chi\left({\chi \over 2}  +  \mu^2 \right) + {i \over 2} \rm{Tr}  \ln G^{-1}
[\chi] \} \]
\begin{equation}
G^{-1}[\chi](x,y)  \equiv \{ \sq  + \chi\ \} \delta (x-y) ,
\label{Ginv}
\end{equation}
Expanding the $\chi$ integral by stationary phase leads to the $1/N$ expansion.
Keeping terms up to quadratic in $\chi$,
perform the ensuing Gaussian integral and then Legendre transform to obtain
the effective action $\Gamma$ we obtain:

\begin{equation}
\Gamma_{eff} [\phi, \chi] =  S_{cl}[\phi, \chi] + 
{\frac{i\hbar}{2}}  \ {\rm Tr} \ln G^{-1}[\chi]+ {\frac{i}{2N}}  \ {\rm Tr} \ln D^{-1}[\phi, \chi],
 \ . 
\label{Seff}
\end{equation}
\bq
D^{-1}[\f, \c](x,y) = -{\frac{1}{\l}}\d^4(x,y) - \f (x)G[\c](x,y) \f(y) + {\frac{i}{2}}G[\c](x,y) G[\c](y,x),
\label{Dinv}.
\eq
The field equation for 
$\phi$ up to order $1/N$ is
\bq
(\sq + \chi (x))\phi (x) - {2 \over N} \int_0^{t_x}dt_y d^3 \vec y \ {\rm Im}
\left[G_>(x,y)D_>(x,y)\right]\phi (y) = 0\ .
\label{fCTP}
\eq
\ba
&\chi(x)& = -\mu^2 + {\l \over 2} \f^2(x) - {i\l\over 4} [G_>(x,x) + G_<(x,x)]\nonumber \\
&+&{\l \over N}\  \ci m \int_0^t dt_1 d^3 \vec x_1 \int_0^{t_1} dt_2 
d^3 \vec x_2 \left[G_>(x,x_1) - G_<(x, x_1)\right] \times\nonumber \\ 
&& \quad \left[\tilde\S_<(x_1,x_2) G_>(x_2,x) - \tilde\S_>(x_1,x_2) 
- G_<(x_2,x) \right]\ ,
\ea
with $\tilde\Sigma$ given by
\bq 
\tilde\Sigma(x_1,x_2) = D(x_1,x_2)\big[iG(x_1,x_2) - \f(x_1)\f(x_2)\big]
\label{sigt}
\eq
Here we have displayed the causal structure that one gets from the CTP Green's
functions.

\section{Summary of Applications}
Because of space limitations, I will just give the references to the applications of these
ideas here. 

Our first studies \cite{back} were on the quantum back reaction problem which completed
Schwinger's classic work on pair production from strong external fields.We
 determined the time evolution of the Electric field from the semiclassical Maxwell 
equation with the current determined by the expectation value of the quantum electromagnetic current resulting from pair production.
 This problem was studied both for a homogeneous plasma
as well as for one that corresponded to boost invariant kinematics appropriate for heavy
ion collisions \cite{kcm}.  

We studied the
dynamics \cite{bib:dcc} of the chiral phase transition in the $O(4)$ sigma model,
both for a boost invariant kinematical situation \cite{dcc} as well as for a radially expanding
plasma of pions \cite{radial}. We showed how to calculate the inclusive distribution of pions as
well as dileptons from the expanding plasma. For the theoretical case of the unbroken $O(4)$ model
we discussed the theoretical question of decoherence, entropy production and determining
the complete density matrix in the mean field approximation \cite{PRL96} \cite{noneq}. 
 
We studied inhomogeneous initial conditions on the fields
as well as solitonic initial conditions \cite{inhom}. Lastly we have been studying in quantum
mechanical scenarios the accuracy of including the $1/N$ corrections to the mean field theory
results. In quantum mechanics, the $1/N$ expansion can only be used for large $N$, with $g/N <
1/70$ because the  order $1/N$ effective potential is not defined for larger $g/N$  at small $x$. 
For these large values of $N$ keeping the non leading corrections to the large $N$ approximation
significantly improves the accuracy \cite{quanN}.

\end{document}